\documentclass{raa}

\usepackage{graphicx,times}             

\begin{document}

   \title{Could bright $\gamma$-ray burst optical transients have been recorded historically?
}

   \volnopage{Vol.0 (200x) No.0, 000--000}      
   \setcounter{page}{1}          

   \author{Richard G. Strom
      \inst{1,2,3,4}
   \and Fuyuan Zhao
      \inst{1}
   \and Chengmin Zhang
      \inst{1}
   }

   \institute{National Astronomical Observatories, Chinese Academy of Sciences,
             Beijing 100012, China \\
        \and
             Netherlands Institute for Radio Astronomy (ASTRON), P.O. Box 2, 7990 AA Dwingeloo, Netherlands; {\it strom@astron.nl}\\
        \and
             Astronomical Institute, University of Amsterdam, Netherlands\\
        \and
            James Cook University, Townsville, Queensland, Australia\\
   }

   \date{Received~~2010 month day; accepted~~2011~~month day}

\abstract{ The brightest optical flash from a $\gamma$-ray burst (GRB) was, briefly, a naked-eye object. Several other GRBs have produced optical transients only slightly fainter. We argue that, based upon the recently accumulated data from hundreds of GRB transients, many such optical events should have been visible to the unaided eye in the course of human history. The most likely repositories of such observations are historical records from the Orient, and we have located and discuss a number of candidates. We also consider the value of such observations, should any very likely ones be uncovered, to modern astrophysics.
}

   \authorrunning{R. G. Strom, F. Y. Zhao \& C. M. Zhang }            
   \titlerunning{Could $\gamma$-ray burst optical transients have been recorded historically?}  

   \maketitle

%
%
\section{Introduction}           
\label{sect:intro}

Optical transients (OTs) associated with GRBs were first observed some 13 years ago (Van Paradijs et al. 1997),
and were the key to understanding the nature of these events, or at least a large fraction of them (Van Paradijs
et al.\ 2000). After attaining their peak brightness, which may be reached in some tens of seconds, most transients
decay rapidly. A quick response by optical telescopes to a GRB detection is therefore essential, but that requires
an equally rapid position determination by the $\gamma$-ray instruments, followed by communication of the information
to optical observatories. The necessary logistics have been available, in several forms, for the past decade, but it
is especially the advent of dedicated GRB observatories like {\it Swift} (Gehrels et al.\ 2004), designed to respond
rapidly (and with its own X-ray and UV-optical telescopes), which has led to the detection of hundreds of OTs.
In the 13 years since the first (pre-{\it Swift}) optical detection --- of GRB~970228 --- well over 300 OTs have been
observed; they have originated in some 40\% of the GRBs discovered over the period. The annual rate of OT
discoveries to the end of 2010 is shown in Fig.\ 1, where the influence of {\it Swift} --- which began operating near the end of 2004 --- is
obvious.

Optically, the brightest $\gamma$-ray burst yet observed was the OT produced by GRB~080319B which peaked at $V=5.3$ mag (Racusin et al. 2008). With a redshift of 0.937, this remarkable event could have been observed with a modest-sized near-IR telescope at $z\simeq17$ (Bloom et al. 2009). In our corner of the universe it was, for half a minute, visible to the unaided eye. It reached its peak brightness just 18 s after the GRB trigger, but by the time the {\it Swift}-UVOT slewed on source, it had faded to $V=7.6$ mag. The GRB~080319B OT onset and peak had been fortuitously captured by several wide-field telescopes which were still monitoring an earlier outburst (GRB 080319A) just $10^\circ$ away. Although the next brightest OT, associated with GRB~990123, was over 3 mag fainter, there is a handful of GRB transients which were not observed until minutes after the trigger and which could have been of similar brightness to GRB~990123. Most of the GRB transients known have been discovered in less than a decade (see Fig.\ 1); it seems inescapable, as we argue below, that there have been significant numbers of naked-eye OTs in the course of human history.

   \begin{figure}
   \centering
   \includegraphics[width=\textwidth, angle=0]{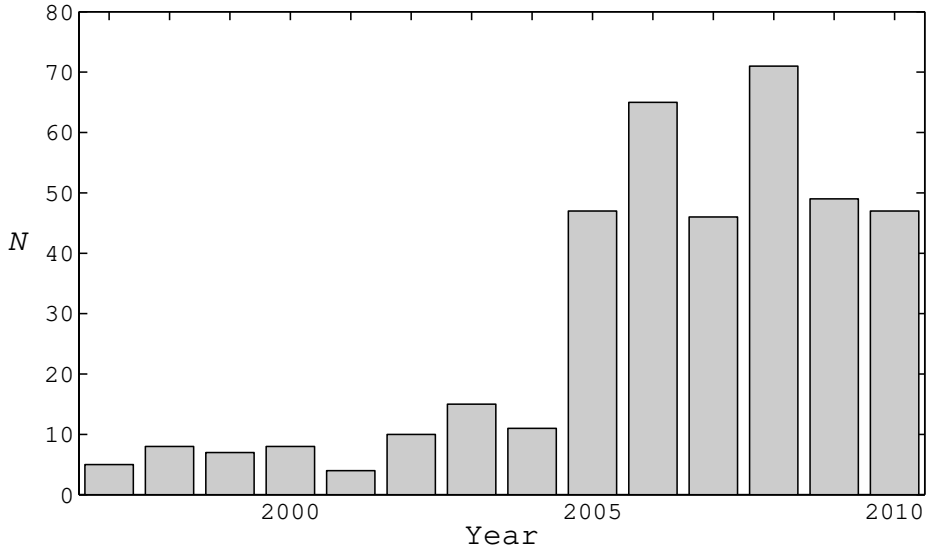}
   \caption{Annual discoveries of OTs associated with GRBs, 1997-2010. }
   \label{Fig:demo1}
   \end{figure}

Transient nighttime events --- in particular, (super)novae, comets, meteors --- have been recorded in a variety of written documents for over 2 millennia. While the bright supernova (SN) of 1006 is mentioned in the annals of a monastery in St.\ Gallen, Switzerland (Clark \& Stephenson 1977; Stephenson \& Green 2002), and the 1066 appearance of Halley's Comet was even embroidered on the Bayeux tapestry (Mardon \& Mardon 2002), no European or other records so far uncovered are the equal of those from East Asia, particularly China. For example, from 240 BC until 1910, every passage of Halley bar one (that of 164 BC) has been described in Chinese historical records (Eddy et al.\ 1989). In this paper we consider what an ancient observer might have seen on the basis of GRB OTs recorded until now, and what description would mark such an event as different from other transient phenomena. We give some examples of records which refer to transients that might have been GRB associated. We hope this preliminary investigation will alert scholars who peruse ancient chronicles to the possible existence of further OT records. We also discuss what the value of such observations, if any can be located, might be to astrophysics.


\section{HOW MANY BRIGHT GRB OTs?}
\label{sect:Obs}

There are several GRB catalogues on the internet keeping track of bursts which have produced an optical counterpart. We have made use of the Berkeley list\footnote{\it lyra.berkeley.edu/grbox/} maintained by D. Perley, which included OTs up to March 2010. To get some idea of the OT brightnesses, Fig.\ 2 is a histogram of the minimum observed magnitude (in bins of $\Delta m=1$) for 330 GRBs in the Berkeley list. It must be emphasized that this is a very heterogeneous collection of OT brightness for several reasons. First, the observations were made in a number of color bands (including unfiltered ``white"), so individual objects could be relatively shifted by 1-2 magnitudes (and certain ``dark" GRBs by even more). There has been little uniformity in the choice of band, and some GRBs were, for a variety of reasons, only observed (or detected) in a single band, making correction to a uniform color problematic. Second, integration times also differ, which will affect the observed brightness of a quickly varying object. The third significant factor is that observation of many GRBs only begins after the light curve has begun to decline. This problem will be discussed further below.

   \begin{figure}
   \centering
   \includegraphics[width=\textwidth, angle=0]{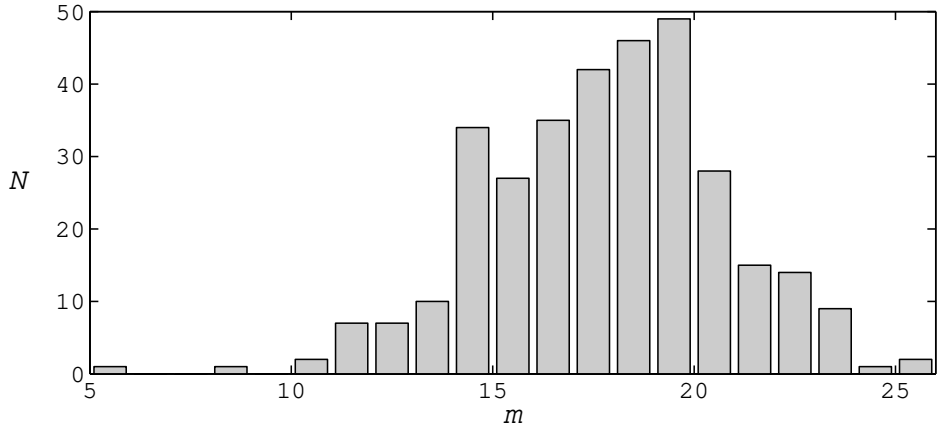}
   \caption{Histogram of GRB OT peak apparent magnitude. }
   \label{Fig:demo1}
   \end{figure}

Despite these shortcomings, the histogram (Fig.\ 2) provides useful information. The distribution is slightly asymmetric, with a peak in the $m=19-20$~mag bin, and a steep decline for $m>20$~mag. Although that aspect does not concern us here, the apparent lack of faint OTs is almost certainly accounted for by the many GRBs undetected in the optical/IR (recall that this group amounts to nearly 60\% of the total). Relevant to our discussion is what happens on the bright side of the distribution, where there is a steady decline to around $m=11$ mag, with a tail extending up to 5.5~mag. As noted above, in many cases the delay between the $\gamma$-ray trigger and the commencement of optical/IR monitoring results in the peak of emission being missed. The slew time for the {\it Swift}-UVOT ranges from $20-70$~s (Gehrels et al.\ 2004), but pointing constraints (Sun, Moon) can delay acquisition further. Moreover, it has been found that for OTs which rise quickly to peak brightness, those which are optically brightest also have the greatest rate of decay (Panaitescu \& Vestrand 2008, Oates et al.\ 2009). Panaitescu \& Vestrand have furthermore found that OTs which peak earlier are more luminous. Hence the result of any delay is to more seriously underestimate the peak brightness of the brightest OTs. It is perhaps no wonder, then, that the brightest OT (of GRB~080319B), which peaked just 18~s after the GRB trigger, was being observed {\it before} the burst occurred (Racusin et al.\ 2008). If we had only the {\it Swift}-UVOT data to go on, its recorded peak would have been over 2~mag fainter. The second brightest transient was also caught before its peak by an extremely rapid follow up (Akerlof et al. 1999). It is worth noting that the bright OT associated with GRB~060117 was considerably brighter than 990123 when first observed at 2 minutes, and they both had nearly the same rate of decay (Jelínek et al.\ 2006; see also Table 1).

%
\begin{table}
\caption[]{ Observed Properties of Bright OTs
}\label{Tab:publ-works}


 \begin{tabular}{lccclr}
  \hline\noalign{\smallskip}
\hphantom{.} GRB & $t_{\rm obs}$ (s) & $t_{\rm peak}$ (s) & OT$_{\rm max}$ (band) & \hphantom{1}$\alpha$ & $V_{\rm peak}$  \\
  \hline\noalign{\smallskip}
990123 & 22 & 47 & 8.86 (V) & 1.80 & 8.9  \\
030329 & 4534 & & 12.35 (nf$^{\rm a}$) & 0.89 & 8.2 \\
050319 & 152$^{\rm b}$ & & 13 (J) & 0.57 & 12.4 \\
051111 & 27 & & 13.0 (nf$^{\rm a}$) & 0.35 & 13.0  \\
060111B & 29.7 & & 13 (nf$^{\rm a}$) & 2.38 & 13  \\
060117 & 128.8 & & 10.12 (R) & 1.73 & 8.8  \\
060418 & 64 & 122 & 10.92 (H) & 1.28 & 11.5  \\
061007 & 137 & & 10.34 (R) & 1.72 & 8.9  \\
061126 & 23 & & 11.9 (R) & 1.48 & 12.0  \\
071003 & 42 & & 12.8 (R) & 1.47 & 12.9  \\
080319B & 0 & 18 & 5.3 (nf$^{\rm a}$) & 2.33 & 5.3  \\
080413A & 20.4 & & 12.8 (nf$^{\rm a}$) & 2.7 & 12.8  \\
080727B & 85 & & 11.8 (H) & 1.08 & 12.4  \\
080810 & 38 & 81 & 12.78 (nf$^{\rm a}$) & 1.22 & 12.8  \\
081121 & 57.2 & & 11.6 (nf$^{\rm a}$) & 3.7 & 11.6  \\
081203A & 134 & 345 & 11.58 (I) & & 11.8  \\
081222 & 28 & 60 & 11.1 (H) & & 11.7  \\
090102 & 43.8 & 88 & 11.78 (H) & 1.5 & 12.4  \\
  \noalign{\smallskip}\hline
\end{tabular} \\
$^{\rm a}$nf = no filter \\
$^{\rm b}$There was a $\gamma$-ray precursor 138 s before the trigger \\
\end{table}

For these reasons, it is more than likely that some of the OTs in Fig.\ 1 actually had significantly brighter peaks. There have been several recent studies of the observed characteristics of OTs. Panaitescu \& Vestrand (2008) have investigated the properties of 30 GRBs with early monitoring of the OT, and find that they can be divided into four classes: fast risers, slow risers, fast decays and plateaus. The largest group (40\% of the total) are the fast decays, GRBs where the peak had already been reached by the time monitoring began. For half of them, the first observation began less than a minute after the trigger, with acquisition times even as short as $20-30$ s. The smaller group of fast risers peaks ``at about 100 s". The OT decay is usually characterized as a power law:   $t^{-\alpha}$. To investigate bright OTs whose peak brightness might have been missed, we consider all GRBs with $m\leq13$ mag in Fig.\ 1. They are listed in Table 1, together with the time of the first observation, $t_{\rm obs}$; the time of the peak in the OT light curve (only if both rise and decay were observed), $t_{\rm peak}$ (times are relative to the GRB trigger); the peak magnitude, OT$_{\rm max}$ (and its observing band); $\alpha$ (for $100\,{\rm s}<t< \rm500\,s$, if available, otherwise from the earliest observation of decay); and a corrected peak visual magnitude, $V_{\rm peak}$, described below. It can be seen that for most of these transients, the peak occurred before the first observation. Their post-peak decline ($\alpha$) is also steep, with a mean of $\langle\alpha\rangle=1.64$ (Oates et al.\ [2009], find $\langle\alpha\rangle=0.88$ for their sample).

In view of the correlation linking large values of $\alpha$ to bright OTs, it seems quite likely that some of the GRBs in Table 1 would have had peak brightnesses significantly greater than the observed values. For those only observed two or more minutes after the GRB trigger, we have used $\alpha$ to determine what the peak magnitude would have been at $t=1$ minute. This choice of 60 s is a compromise between the ``fast risers" and ``fast decays" (Panaitescu \& Vestrand 2008), and we feel it is a conservative choice. In addition, the observed brightness (when not made in the visual band) has been corrected to V-band, based upon the typical power-law color index of 0.5 ($I\propto\lambda^{0.5}$) found by Van Paradijs et al.\ (2000), and also observed in recent GRBs including 080319B (Bloom et al.\ 2009). This gives us $V_{\rm peak}$ values (Table 1), resulting in a change to the bright tail of the distribution (Fig.\ 2), as shown in Fig.\ 3.

   \begin{figure}
   \centering
   \includegraphics[width=\textwidth, angle=0]{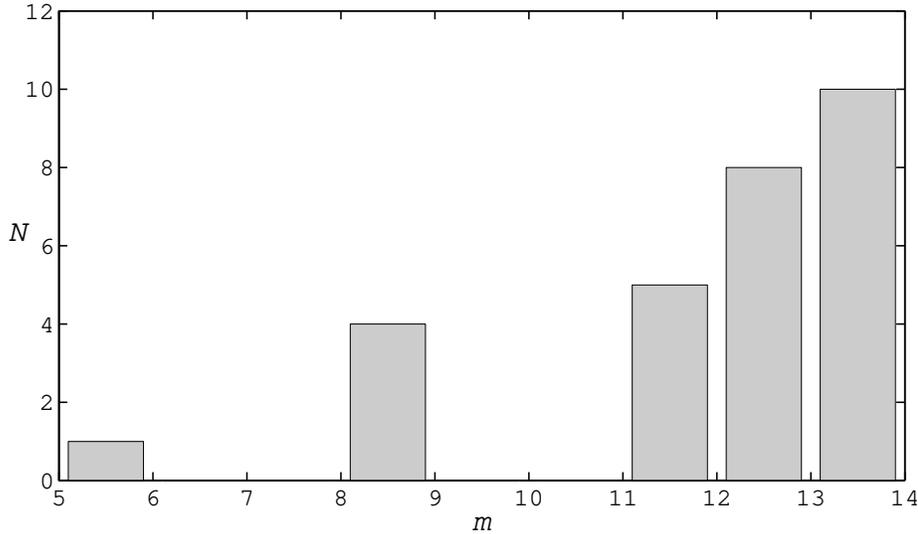}
   \caption{Bright end of histogram (Fig.\ 2) after recalculation of $V\rm_{peak}$. }
   \label{Fig:demo1}
   \end{figure}

We have fit the numbers of sources in the declining part of the bright tail (for $m\leq17$) as a function of magnitude, and find a very good power-law relationship. Extrapolating the correlation to lower $m$, we find that $V=3$~mag OTs will occur one-fifth as often as those of $V=6$~mag, while $V=0$~mag will be one-twentieth as frequent. The OTs of GRBs have been investigated for about a decade (the period of rapid follow-up is actually closer to 5 years, Fig.\ 1), so for 5.5~mag objects we estimate at least 10 per century, and about 2 per century for 3~mag. A handful of $\sim0$~mag OTs might be expected in a millennium.

\section{HISTORICAL RECORDS}
\label{sect:data}

Bright nighttime transients have been recorded by several civilizations for thousands of years. Particularly relevant here are the ``new" stars which today are associated with supernovae (SNe). A naked-eye OT has much in common with a bright (super)nova: both are star-like and have fixed positions. Their appearance is unpredictable in time and location. However, a Galactic object, as most SNe would be, should have a low galactic latitude (a SN in the Magellanic Clouds being an obvious exception); GRBs, being overwhelmingly extragalactic, are uniformly distributed over the sky. The distribution of bright novae, which are also galactic, reaches to higher latitudes because of their smaller average distances (and their fainter intrinsic brightness means that only nearby ones are naked-eye objects; see Clark \& Stephenson 1977).

The main difference between a GRB visual event and the other stellar transients is one of timescale. The OT associated with a GRB will be visible for a matter of minutes, in a few cases hours, while a supernova of the same magnitude should be observable for days or longer. Even a fast nova would only disappear after a few days. As an example let us consider the OT of GRB~080319B. Bloom et al.\ (2009) have used the comprehensive observations of this GRB to construct the OT light curve for a wide range of redshifts (see their Fig.\ 9). If it had been closer ($z\simeq0.4$), it would have been brighter than 6 mag (and so visible to the unaided eye) for about a minute. A type Ia SN of the same magnitude could have been observed for 44 days (Branch \& Tammann 1992). Even a very fast nova like V603 Aql would have been visible for 6 days (Wyse 1940). So the type of historical record we are looking for would mention the sudden appearance of a star at a random location in the sky, and its disappearance shortly thereafter. One might wonder whether such a brief appearance would have been noticed (and recorded) by ancient observers. In East Asia the answer is that it most likely would have been. Chinese records are replete with observations of meteors, with timescales of a second or less, so fast transients would not have been excluded because of their short duration.

Most meteors are seen to move, however, so their appearance is noticeably different from a GRB OT, which is perfectly stationary. Would this fact make the observation of an OT by an experienced observer less likely? There is relevant information on this question from 19$^{\rm th}$ and 20$^{\rm th}$ Century visual observers. An investigation of flashes which were suggested to have an astronomical origin (Katz et al.\ 1986) provides, in addition to evidence against the astronomical interpretation (Schaefer et al.\ 1987), data on the observability of brief star-like flashes covering a range of apparent magnitudes. Observations made in 1984-5 by experienced meteor observers (most of whose ``sightings were accidental and not part of an organized survey") recorded 26 flashes, half of which had estimated magnitudes ranging from +1 to +6 (Katz et al.\ 1986) and durations $0.2\pm0.1$ s to $1.5\pm0.7$ s. This demonstrates that even faint star-like events of extremely short duration can be detected visually (and the OT flashes we consider here would be visible not for a fraction of a second, but for 10s of seconds or more).

Schaefer et al.\ (1987), who conclude that most of the flashes seen in recent years were caused by satellite ``glints", report on control studies which visually searched for, found and recorded brief flashes such as those catalogued by Katz et al. The experienced observer Brian Warner discusses both his own experience, and refers to visual observations of the brief brightening of $\zeta$ Lyrae in 1850 (Warner 1986). In the latter case, the remarkably keen-sighted variable star observer Eduard Heis saw how $\zeta$ Lyrae (which he used as a comparison star in naked eye observations of $\beta$ Lyr) `` `became for a {\it moment very bright}, and then again faint' (`$\beta$ Lyrae wurde einen {\it Moment sehr hell} und hierauf wieder dunkel')" (Hagen 1903). The normal magnitude of $\zeta$ Lyrae is 4.5 and Hagen estimates that it ``probably did not reach the brightness of" $V=1$. Warner (1986) himself, while visually scrutinizing $\mu$ and $\nu$ Cen, observed a ``brief flash of about second magnitude" which he quickly ascertained to be sunlight reflected from a satellite. Oriental astronomers one or two millennia ago are hardly likely to have been less capable of observing and recording brief flashes than the visual observers of the last 200 years. Having said that, in fact the instant of an OT's brightening does not itself have to be seen (though its sudden appearance may help draw the observer's attention). The crucial point is that someone must examine the sky where the OT has appeared and recognize that a new object is present, which is just the way new comets and stars have been discovered for several millennia.

In addition to the temporal behavior, we have to consider the faintest visual magnitude an OT could have yet still be noticed by an observer. Clark \& Stephenson (1977) have investigated the matter. They note that the variable star Mira Ceti was seldom if ever recorded in Chinese annals, though its peak magnitude ranges from +2 to +4 mag (at minimum it is $\sim+9$ mag) and it lies in a region relatively free of other bright stars. On this basis they reckon that a transient would have to be +3 mag or brighter to have a good chance of being discovered. On the other hand, Strom (2011) has investigated the completeness of Oriental records and argues on the basis of the numbers recorded that comets as faint as +4.5 mag must have been discovered. In the following we will take a peak magnitude of +3.5 for our examples.

We have made a limited search in a catalogue of astronomical texts compiled from a wide range of historical sources (including, but not limited to, 24 imperial histories and the Qing draft history) by a workgroup of scholars (Beijing Astronomical Observatory [BAO] 1988); in the following discussion we will refer to extracts from this catalogue by giving the relevant page number. We have found nine records of interest here.

\begin{enumerate}

\item{On a day corresponding to 1059 December 3 (BAO, p.\ 680): ``A star appeared beside {\it Tiancang}." {\it Tiancang} is an asterism in Cetus (Sun \& Kistemaker 1997). The approximate location indicated for the OT is, $\rm RA=0^h$, Dec $=-15^\circ$ (positions, which are very rough, are given for the present epoch). This record (listed under meteors) provides the minimum information required: a date and a location, but does not say if/when the star disappeared.}

\item{On 1057 October 3 (BAO, p.\ 677): ``A star appeared at the side of the {\it Nanhe} stars." {\it Nanhe} consists of three stars, principally $\alpha$ and $\beta$ Canis Minoris (Sun \& Kistemaker 1997). The approximate position corresponds to, $\rm RA=07^h\,\rm30^m$, Dec $=+5^\circ$. The passage is similar to \#1, and is also found under meteors; here again nothing is said about its disappearance. There may be some connection with the following entry.}

\item{On 1430 September 9 (BAO, p.\ 377): ``Nighttime, a guest star appeared at the side of {\it Nanhe}, like a bullet in size, dark color [literally, ``blue-black"], altogether 26 days [then] extinguished." On the basis of our previous discussion, 26 days is too long for a GRB, and in view of \#2, this may have been a recurrent nova. For a discussion of the term ``like a bullet" and similar comparisons, see Strom (2008).}

\item{On 1431 January 4 (BAO, p.\ 377): ``There is a star like a bullet, seen at the side of {\it Jiuyou}, smooth and glossy pale yellow, 15 days then invisible." {\it Jiuyou} is an asterism in or near Eridanus. The approximate position would be, $\rm RA=05^h$, Dec $=-10^\circ$. This entry has all the information needed: date, location and period of visibility, but 15 days is also probably too long for a GRB.}

\item{On 1175 August 10 (BAO, p.\ 379): ``There is a {\it bei} star in the northwest direction, in the presence of and outside {\it Ziwei Yuan}, upper part of {\it Qigong}, small as Mars, awe-inspiring disheveled {\it bei}, until on a {\it bingwu} [day it] began to disappear." {\it Ziwei} is the celestial palace wall around the North Polar Region, encompassing the constellations Draco, Ursa Major, Ursa Minor and Camelopardalis (Sun \& Kistemaker 1997). {\it Qigong} is an asterism straddling Draco and Bootes; from the description the {\it bei} star was probably in Draco, near $\rm RA=16^h$, Dec $=+55^\circ$. A {\it bingwu} day is the $\rm43^{rd}$ in the sexagenary cycle, which was five days after the discovery date (a {\it xinchou} day). This and the following entry are listed as uncertain new stars. A {\it bei} star ({\it bei} [or {\it bo}] {\it xing}, sometimes translated as ``sparkling star"; here the order is reversed: {\it xing bei}) is one of the technical terms often applied to a comet which ``sends out its rays evenly in all directions", perhaps one with its tail directed away from the observer (Ho 1962). The use of {\it bei} and the description disheveled may suggest that the ``star" was extended. In any event the object was probably visible for several weeks, as the next entry appears to be a continuation.}

\item{On 1175 August 11 (BAO, p.\ 379): ``Nighttime, {\it bei} star seen in western direction, [after] 27 days and nights disappears and subsides." Taken with the previous entry, the duration of 28 days is too long for a GRB.}

\item{On 1688 November 2 (BAO, p.\ 380): ``Unusual star seen in {\it Kui}, color white, altogether 3 nights." {\it Kui} is an asterism in Andromeda, and the fifteenth lunar mansion (Sun \& Kistemaker 1997). The approximate location is, $\rm RA=1^h$, Dec $=+30^\circ$. Except for the location, color and duration, this description is exactly like the following one two years later.}

\item{On 1690 September 29 (BAO, p.\ 380): ``Unusual star seen in {\it Qi}, color yellow, altogether 2 nights." {\it Qi} is a small asterism in Aquarius (Sun \& Kistemaker 1997). The approximate location would be, $\rm RA=22^h$, Dec $=-7^\circ$. Could a GRB OT have been seen for as long as two or three nights? And if so, could it be distinguished from a nova? GRB~060614, described as a plateau OT (Panaitescu \& Vestrand 2008), had a brightness which would have remained between $+3.5$ and $+6$~mag for 2.3~d, so it is possible. The most rapid nova known is U Sco (Kato 2002) which declines at a rate of 0.6~mag~day$^{-1}$. At its 1999 outburst, U Sco's brightness was within 2.5~mag of its peak for 3.9~d (Munari et al.\ 1999).}

\item{On 1855 August 16 (BAO, p.\ 1081): ``There is a star of red color, large as a {\it yu}, seen in the southeast, for about 7 minutes." A {\it yu} is a basin or wide-mouthed jar. Containers are often used as comparison objects when describing transients (Strom 2008). The time given in the original text is ``half a quarter". In terms of timescale, this is the best candidate. Unfortunately, the position given is too vague to carry the investigation further.}

\end{enumerate}

The Chinese records we have consulted are more complete for some astronomical phenomena than others. For comets and supernovae, it is estimated that Chinese annals contain about half of the naked-eye objects which appeared (Strom 2011). In the case of a bright comet like Halley, 28 of its 29 appearances from 240 BC were chronicled. Yet the record for sunspots is woefully incomplete, with less than 1\% of those which could have been seen actually recorded somewhere (Eddy et al.\ 1989). But sunspots are a daytime phenomenon, and hopefully nighttime transients would fare as well as comets do. The most complete record is likely to be found in the Korean {\it Seungjeongwon Ilgi} (Stephenson 2011), a chronicle covering almost the entire period 1623-1894. Whereas the Chinese dynastic histories include only a selection of celestial observations, based upon (presumably more extensive) original source material, the {\it Ilgi} is an essentially daily record of ``all official business" of the royal court, as well as ``significant events", including astronomical ones.

Stephenson discusses a number of examples, in particular one somewhat relevant to GRB transients: the disappearance of a star. On 3 October 1625, the {\it Ilgi} records, ``At night in the first watch, the star {\it Ziwei Yuan Tianyi} was not seen" (Stephenson 2011). Then for most of the following 22 days the continued absence of {\it Tianyi} is noted. Thereafter, the star is not mentioned further, nor is its reappearance recorded in the extant text to mid-1628. {\it Tianyi} is believed to be 10 Dra, an M3 giant. Its peculiar behavior (if the identification is correct) is unexplained. But such a detailed record is what is required in a search for rapid transients. Unfortunately, locating candidate events in the text will be a massive undertaking: the {\it Ilgi} is estimated to contain some 150 million characters; in facsimile copies the text fills about 100 large volumes (Stephenson 2011).

\section{DISCUSSION}
\label{sect:analysis}

It is probably difficult to countenance the possibility that ``crude" naked-eye observations made in ancient times could add significantly to our understanding of GRBs. However, although the human eye has its limitations when it comes to sensitivity, it can pick up sudden changes over a wide field of view. Even today, useful observations of meteor showers are made with the unaided eye (usually by experienced amateurs). A major limitation is that what an ancient observer chose to record (or not) will probably differ from what today's astrophysicist would regard as important.

An ideal record would mention the sudden appearance of a ``guest star" ({\it ke xing}), its disappearance a short time later and the fact that it was motionless. Some description of brightness would be nice, but more important is an indication of its location. Because of their rareness, all GRBs have been discovered in external (and usually distant) galaxies. A GRB in the Milky Way would be improbable, though of immense interest. It would probably go off in the inner Galaxy, where the highest concentration of stars is. At a distance of 8~kpc, GRB~080319B would have $m=-24.08$~mag. But the fact that the extinction to the galactic center is some $A_v=28$~mag (Oort 1977) means that an observer might see a $+4$~mag peak OT. A GRB transient would be more likely to originate in an external galaxy, which would only be identifiable with a good position. One would have to hope that some bright OTs have taken place in nearby galaxies.

We know nothing about the remnants which GRBs produce. The discovery of an historical sighting would offer the possibility of examining its remnant after hundreds or thousands of years. While one expects to see something like a supernova remnant, they come in a variety of shapes, notably the filled-center (Crab Nebula) type, and shell or ring-like. Even if a remnant cannot be unambiguously located, the identification of an ensemble of visual OT candidates would provide statistics on the rare, optically bright GRBs.

\begin{acknowledgements}
The research of RGS has been supported by a Chinese Academy of Sciences Visiting Professorship for Senior International Scientists. Grant Number: 2009J2-1.
\end{acknowledgements}

\label{lastpage}

\end{document}